\documentclass[preprint2]{aastex}
\begin{document}

\title{Activity and Kinematics of Ultracool Dwarfs Including An Amazing Flare 
Observation}

\author{Sarah J. Schmidt,\altaffilmark{1,2,3,4} 
Kelle L. Cruz,\altaffilmark{1,4,5} Bethany J. Bongiorno,\altaffilmark{1,2} 
James Liebert,\altaffilmark{4,6} and I. Neill Reid\altaffilmark{4,7}}

\altaffiltext{1}{Department of Astrophysics, American Museum of
Natural History, New York, NY}
\altaffiltext{2}{Department of Physics \& Astronomy, Barnard College, 
Columbia University, New York, NY} 
\altaffiltext{3}{Department of Astronomy, University of Washington, 
Seattle, WA}
\email{sjschmidt@astro.washington.edu}
\altaffiltext{4}{Visiting astronomer, Kitt Peak National Observatory and/or 
Cerro Tololo Inter-American Observatory, National Optical Astronomy 
Observatories, which are operated by the Association of Universities for 
Research in Astronomy, under contract with the National Science Foundation.}
\altaffiltext{5}{NSF Astronomy and Astrophysics Postdoctoral Fellow}
\altaffiltext{6}{Department of Astronomy and Steward Observatory,
University of Arizona, Tucson, AZ}
\altaffiltext{7}{Space Telescope Science Institute, Baltimore, MD}

\begin{abstract}
We present the activity and kinematics of a representative volume-limited 
(20~pc) sample of 152 late-M and L dwarfs (M7--L8) photometrically selected 
from the Two Micron All-Sky Survey (2MASS). Using new proper motion 
measurements and spectrophotometric distance estimates, we calculate 
tangential velocities. The sample has a mean tangential velocity of 
$\langle V_{tan} \rangle = 31.5~km~s^{-1}$, a velocity dispersion of 
$\sigma_{tan} = 20.7~km~s^{-1}$, and a maximum tangential velocity of 
V$_{tan} = 138.8~km~s^{-1}$. These kinematic results are in excellent 
agreement with previous studies of ultracool dwarfs in the local solar 
neighborhood. H$\alpha$ emission, an indicator of chromospheric activity, was 
detected in 63 of 81 late-M dwarfs and 16 of 69 L dwarfs examined. We find a 
lack of correlation between activity strength, measured by 
$log(F_{H\alpha}/F_{bol})$, and V$_{tan}$, though velocity distributions 
suggest that the active dwarfs in our sample are slightly younger than the 
inactive dwarfs. Consistent with previous studies of activity in ultracool 
dwarfs, we find that the fraction of H$\alpha$ emitting objects per spectral 
type peaks at spectral type M7 and declines through mid-L dwarfs. Activity 
strength is similarly correlated with spectral type for spectral types later 
than M7. Eleven dwarfs out of 150 show evidence of variability, ranging from 
small fluctuations to large flare events. We estimate a flare cycle of 
$\sim$5\% for late-M dwarfs and $\sim$2\% for L dwarfs. Observations of 
strong, variable activity on the L1 dwarf 2MASS~J10224821+5825453 and an 
amazing flare event on the the M7 dwarf 2MASS~J1028404$-$143843 are discussed.
\end{abstract}

\keywords{Galaxy: stellar content --- solar neighborhood 
--- stars: activity --- stars: flare
--- stars: late-type stars: low-mass, brown dwarfs ---  
stars: individual (2MASS~J10224821+5825453, 2MASS~J1028404$-$143843)}

\section{Introduction}
\label{sec:intro}
The identification of large numbers of early-to-mid M dwarfs has enabled 
studies focused on their activity and kinematics 
\citep{Reid95,PMSU2,Bochanski05} and recent work has extended these 
investigations to late-M and early-L dwarfs \citep{NN,West04,West06} but faint 
magnitudes and small numbers have prevented a thorough study of the activity 
and kinematics of L dwarfs. In order to fill that gap, we analyze a sample 
of ultracool dwarfs (spectral types M7--L8) photometrically selected from the 
the Two Micron All Sky Survey \citep[2MASS]{2MASS}. The Two Micron Proper 
Motion sample (2MUPM sample) is a combination of a complete sample selected 
from the 2MASS Second Incremental Data Release \citep{Paper5,Paper9} and the 
spectroscopically confirmed portion of a sample constructed from the All-Sky 
Data Release \citep{Paper10}. This comprehensive survey allows us to 
investigate the activity and kinematics of cooler, fainter dwarfs in a sample 
with well documented properties and uncertainties.

We measured proper motions using the images provided by the Digitized Sky 
Survey (hereafter DSS), which combines data from the Palomar Observatory Sky 
Survey \citep[POSS]{POSS} and the UK Schmidt survey \citep{UKSchmidt}. 
Together, DSS images and 2MASS coordinates and images have enabled long 
baseline proper motion measurements for all but the 12 faintest objects, and 
the inclusion of new data has allowed us to obtain proper motions for all but 
seven dwarfs. We combine our proper motions with spectrophotometric distance 
estimates to obtain tangential velocities for our sample of 152 ultracool 
dwarfs. Kinematics allow us to investigate the age distribution of our sample; 
the ages of L dwarfs are of particular interest because the spectral class 
encompasses both stellar and substellar objects \citep{BRev01}.

Kinematics can also be combined with H$\alpha$ measurements to test the 
validity of an age activity relationship for ultracool dwarfs. The power law 
relationship between activity and age in main sequence stars 
\citep{Skumanich72,Soderblom91} breaks down during the M spectral class 
\citep{Silvestri05,NN}. Kinematic studies show that active M dwarfs tend to be 
younger than inactive M dwarfs, but it is unknown whether this is due to a 
sharp cut-off in activity as dwarfs age, or a steady decline 
\citep{West04,West06}.

Activity is more closely related to spectral type in ultracool dwarfs. For 
M1 to M7 dwarfs, the fraction of active objects increases with later spectral 
type while the strength of H$\alpha$ emission remains relatively constant, 
though with scatter \citep{PMSU2,West04,Bochanski05}. For M7 to late-L dwarfs, 
H$\alpha$ emission becomes increasingly less common with later spectral type 
as the activity strength decreases \citep{NN,West04}. There is evidence that 
this relationship may extend into the T spectral class \citep{Burgasser02} 

The mechanisms producing flare events and strong H$\alpha$ emission in 
ultracool dwarfs are not completely understood. Current estimates of the 
M dwarfs flare flare rate are approximately 7\% \citep{NN,Reid99}. 
Only a few L dwarfs have been observed with strong H$\alpha$ emission and 
little is known about their flare properties 
\citep{Liebert99,Liebert03,Hall02}.

We present proper motions and H$\alpha$ measurements of a sample of 152 
ultracool dwarfs with spectral types from M7--L8. In \S~\ref{sec:sample}, we 
discuss the completeness and properties of our sample. Details of proper 
motion measurements and the examination of spectral features are contained in 
\S~\ref{sec:measure}. We discuss the kinematics in \S~\ref{sec:KineResults} 
and in \S~\ref{sec:ActResults} we examine the activity properties of the 
sample and investigate possible age relations. Variability is discussed 
in \S~\ref{sec:FlareResults}, including an strong variability on the L1 dwarf 
2MASS~J10224821+5825453 and an amazing flare on the M7 dwarf 
2MASS~J1028404$-$143843.

\section{The Sample}
\label{sec:sample}
To construct the 2MUPM sample of late-M and L dwarfs discussed in this paper, 
we have combined objects selected from both the 2MASS Second Incremental Data 
Release and the All-Sky Data Release. The dwarfs selected out of the second 
release are from \citet[hereafter Paper V]{Paper5} and \citet[hereafter Paper 
IX]{Paper9}. The dwarfs selected out of the all-sky release are from 
\citet[hereafter Paper X]{Paper10}. Those papers describe how extensive 
follow up spectroscopy was obtained for candidate objects from both samples. 
Spectroscopy has been completed for the objects selected from the second 
release, but more observations are needed to finish confirmation and spectral 
typing for the objects selected from the all-sky release.

Though we summarize here, more details on sample selection, spectroscopy, 
spectral types, and spectrophotometric distance estimates can be found in 
Papers V, IX, and X. Spectral types were assigned by visual comparison of each 
spectra to spectral standards. There is an uncertainty of $\pm$0.5 spectral 
subtype for most dwarfs, and whole number types are favored over half types. 
Spectrophotometric distance estimates were derived from a relation between 
spectral type and absolute \textit{J} magnitude. 

The subset of the sample selected from the 2MASS second release is complete 
for spectral types M9--L6 within a distance limit of 20~pc, but we include 
types M7--L8 in the 2MUPM sample. The color selection ($J-K>1$) excludes the 
bluest M7 and M8 dwarfs, and magnitude limits exclude the most distant L7 and 
L8 dwarfs. Despite this incompleteness, there is no evidence that the near 
infrared selection criteria used to create the sample is correlated with 
activity or kinematics. The 2MASS second release sample comprises only 35 L 
dwarfs within 20 parsecs. In order to increase the sample size, we have 
included spectroscopically classified dwarfs within 20 pc from the all-sky 
sample, which adds 37 L dwarfs to the 2MUPM sample. As 
described above, we have not yet completed observations of the faintest 
objects from latter dataset; however, those omissions should not bias the 
activity or kinematic distributions of the sample as a whole.

\begin{figure}
\plotone{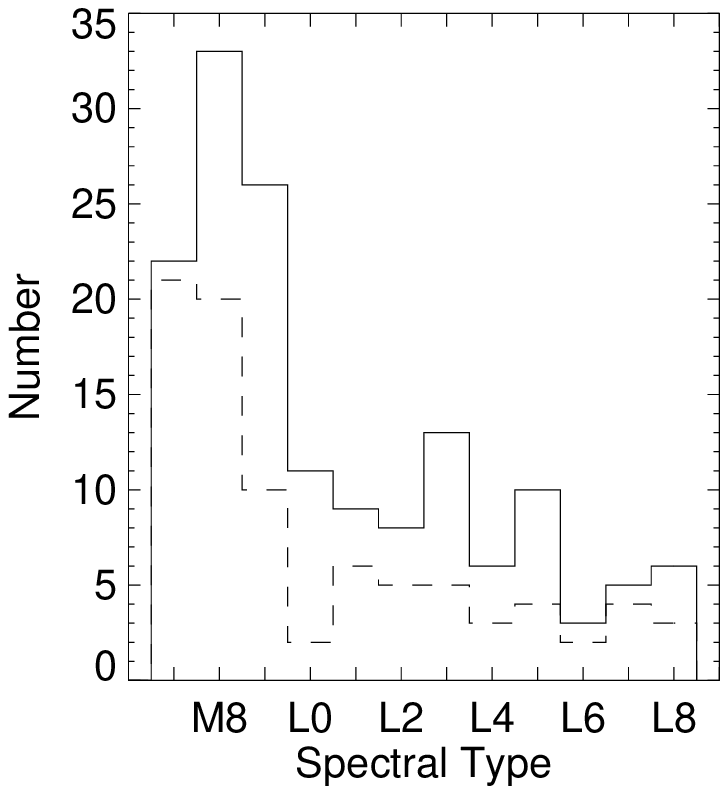}
\caption[Sample Histogram]{Spectral type distribution of the 2MUPM sample. The 
objects selected from the 2MASS second release \textit{(dashed line)} and the 
objects selected from both the second and the all-sky release 
\textit{(solid line)} are shown. A K-S test for types M8--L8 (see 
\S~\ref{sec:sample}) indicates no difference between the two spectral 
type distributions at the 90\% significance level.}\label{fig:sample}
\end{figure}

The sample selected from the 2MASS second release is the basis of the 
luminosity function of ultracool dwarfs. Excepting a small number of dwarfs 
excluded because we could not measure proper motions, that portion of the 
2MUPM sample has the same completeness discussed in Paper IX. While 
observations of the sample selected from the 2MASS all-sky release have not 
been completed, it was created with similar selection criteria. 
Figure~\ref{fig:sample} shows the number of objects per spectral type included 
from the second release selection and the all-sky selection. Excepting M7 
dwarfs, which were purposely excluded from the all-sky selected sample, a K-S 
test indicates no difference between the two spectral type distributions at 
the 90\% significance level. 

The 2MASS-selected 20~parsec sample is currently composed of 170 dwarfs with 
spectral types from M7 to L8. The 2MUPM sample includes 11 multiple systems 
without resolved spectroscopy. The spectral types of both the primary and the 
secondary in these systems are estimated to be between M7 and L8. Because the 
lack of resolved spectroscopy does not allow us to check each object for 
H$\alpha$ emission, we consider the system as a single object with the 
spectral type of the primary. This excludes 11 low mass companion objects from 
our analysis. Treating binaries as single object also affects the relative 
strength of H$\alpha$ emission, which is discussed in \S~\ref{sec:ActResults}. 
\object{LHS 1070} is a resolved triple system (M5.5, M8.5, and L0) where the 
primary is excluded by our spectral type criterion, but the M8.5 and L0 are 
part of the 2MUPM sample \citep{Leinert00}.

Six objects are excluded from the 2MUPM sample because their optical 
magnitudes were too faint to obtain or measure a proper motion; one object was 
too close to another source to measure an accurate position. The proper motion 
measurements are discussed further in \S~\ref{sec:KineMeas}. With the 
exclusion of 7 objects without proper motions and 11 low mass companion 
objects without resolved spectroscopy, the remaining 152 objects (81~M and 
71~L) comprise the 2MUPM sample analyzed in this paper. 

\section{Measurements}
\label{sec:measure}
\subsection{Proper Motions and Tangential Velocities}
\label{sec:KineMeas}
When combined, 2MASS and DSS enable accurate, long baseline proper motion 
measurements for most 2MUPM systems---DSS images are from as early as 1950, 
while 2MASS coordinates and infrared images are from between 1997 and 2002. 
From the 159 2MASS selected M7--L8 dwarfs within 20 pc, we measured proper 
motions for 81 of 82 M dwarfs and 66 of 77 L dwarfs using DSS data as the 
first epoch and 2MASS coordinates as the second. The proper motion 
measurements adopted in this paper are a balance between the longest possible 
baseline and the best quality image. For 20\% of the sample, POSS~I data 
(baseline of 40--50 years) was used. Ten percent of the proper motions were 
measured using UK Schmidt data (16--24 years). The remaining proper motions 
were measured mainly from POSS~II data with baselines of 2--16 years (20\% 
with baselines $<$5 years). 

While many of these objects were visible in multiple epochs of DSS data, 
eleven late-L dwarfs (2 L6, 5 L7, 4 L8) are not visible in any DSS band due to 
their faint optical magnitudes. One proper motion is available in existing 
literature; we use the measurement from \citet{Vrba04} for 
2MASS~J01075242+0041563. Data were obtained to measure proper motions of four 
of the remaining faint L dwarfs. The proper motions for 2MASSI~J1043075+222523 
and 2MASSI~J2325453+425148 were measured with 2MASS data as the first epoch 
and deep, high resolution Gemini acquisition images from our 2004B program 
(Program ID: GN-2004B-Q-10) as the second epoch. For 2MASS~J02572581$-$3105523 
and 2MASSI~J0439010$-$235308, we used data from the CPAPIR camera on the CTIO 
1.5m to provide second epoch astrometry. We exclude the remaining 6 L dwarfs 
(1 L6, 5 L7) from the 2MUPM sample. We also exclude one M8.5 dwarf because it 
is too close to a background star to measure an accurate position.

For each object, multiple band DSS images were obtained from the Canadian 
Astronomy Data Centre or, in four cases, another image was acquired of the 
target and surrounding field. Additionally, a list of 2MASS coordinates for 
nearby reference stars was obtained from the NASA/IPAC Infrared Science 
Archive using the GATOR interface. Proper motions were then measured using a 
custom IDL code. For each object, a transformation matrix between DSS image 
x,y coordinates and 2MASS frame RA,dec is calculated using the positions of 
reference stars surrounding the proper motion source. The position of the 
source in DSS image x,y coordinates is then translated into 2MASS frame RA,dec 
using the transformation matrix. The proper motion is solved for by comparing 
the translated first epoch coordinates with the 2MASS coordinates. 
Uncertainties in the proper motion measurements are calculated from the 
residuals between the transformed coordinates and the 2MASS coordinates for 
the reference stars; they are inversely dependent on the length of the 
baseline.

The tangential component of space velocity (V$_{tan}$) is the product of 
proper motion and distance. Our distances are calculated from the spectral 
type/M$_J$ relation found in Paper V. Proper motions, distances, and derived 
tangential velocities are given in Table~\ref{tab:all}. The V$_{tan}$ 
distribution has a mean of $\langle V_{tan} \rangle = 31.3~km~s^{-1}$ and a 
dispersion of $\sigma_{tan} = 20.8~km~s^{-1}$. Kinematic results are discussed 
in more detail in \S~\ref{sec:KineResults}. Most objects in our sample have 
V$_{tan}$ less than 100~km~s$^{-1}$ and the two kinematic outliers 
(V$_{tan} > 100~km~s^{-1}$) are discussed in \S~\ref{sec:FastObjects}.

It was not possible to measure complete UVW kinematics for the 2MUPM sample. 
The spectra were not obtained with the goal of measuring radial velocities; 
they are from a variety of telescopes and conditions. Radial velocities 
require more consistent data. Additionally, the spectra for fainter dwarfs 
do not have sufficient signal-to-noise to measure radial velocities so the 
sample would be incomplete in later spectral types. While the lack of complete 
space motions does effect our analysis of individual objects, it should not 
drastically effect the kinematics of the population \citep{Silvestri02}.

\begin{figure}
\plotone{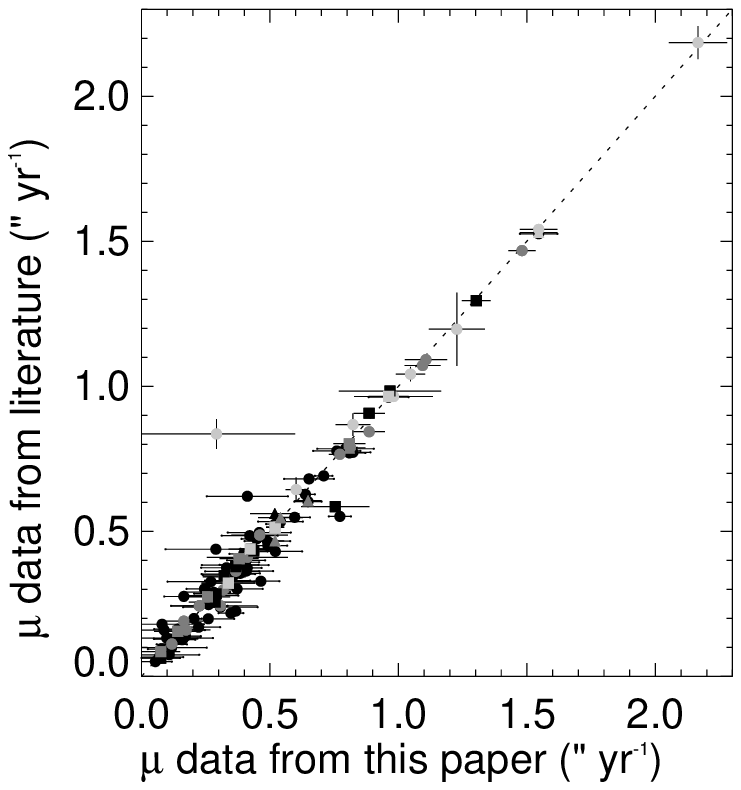}
\caption[Proper motion comparison]{Proper motions from this paper vs. 
previously published measurements. The dotted line represents perfect 
agreement between the measurements. Data is plotted from the USNO-B catalog 
\textit{(black circles)}, \citet[\textit{dark grey circles}]{NN},
\citet[\textit{light grey circles}]{Deacon05}, 
\citet[\textit{black squares}]{Dahn02}, 
\citet[\textit{dark grey squares}]{Tinney95}, 
\citet[\textit{light grey squares}]{Tinney96}, 
\citet[\textit{black triangles}]{TK03}, 
\citet[\textit{dark grey triangles}]{Monet92}, and 
\citet[\textit{light grey triangles}]{Jao05}.}\label{fig:PM}
\end{figure}

Proper motions were previously published for 95 of the 152 objects in the 
2MUPM sample. We compared these measurements with ours to perform a check on 
our method of measuring proper motions. For 44 objects, measurements were made 
by either astrometric studies \citep{Monet92,Tinney95,Tinney96,Dahn02,Vrba04} 
or kinematic studies \citep{NN,Deacon05}. Additionally, 80 of the 152 objects, 
proper motions were published in the USNO-B astrometric catalog 
\citep{USNO-B}. Our measurements are plotted against previous measurements in 
Figure~\ref{fig:PM}.

The USNO-B measurements are not completely consistent with ours. Seventy of 80 
objects agree within 0.1~\arcsec~yr$^{-1}$. Of the remaining 10, five agree 
within 0.2~\arcsec~yr$^{-1}$, and five did not agree (within 
$<$0.2~\arcsec~yr$^{-1}$; these are not included in Figure~\ref{fig:PM}). 
The automated methods used to measure proper motions for the USNO-B catalog 
are prone to mismatches; we prefer our proper motions as each measurement has 
been verified by eye.

Comparison with other works shows good agreement; our measurements are within 
0.1~\arcsec~yr$^{-1}$ of previous $\mu_{tot}$ and 10\degr~of previously 
measured position angle. The one exception is SIPS 1936$-$5502, which we 
measure at $\mu_{tot}=0.29\pm0.31\arcsec~yr^{-1}$, PA=131\degr~(with a three 
year baseline) and \citet{Deacon05} measure at 
$\mu_{tot}=0.84\pm0.05\arcsec~yr^{-1}$, PA=134\degr. It is possible that our 
measurement is not entirely correct due to the short baseline, but we retain 
the object in our sample as it would be included in the absence of a 
comparison measurement. Generally, the proper motions found by parallax 
studies are more accurate, but we use our own measurements for consistency. 

Twenty-one of the 60 objects analyzed by \citet{NN} were re-observed as part 
of the 2MUPM sample. The proper motions show agreement within 
0.06~\arcsec~yr$^{-1}$ (within our uncertainties) but the V$_{tan}$ comparison 
is less consistent. The distance estimates in \citet{NN} are based on $J-K_S$ 
color while our distances are derived from a spectral type/M$_J$ relation. 
The latter method yields improved distances (accurate to $\sim$10\%) and thus 
more precise kinematics. For example, the maximum V$_{tan}$ of 141~km~s$^{-1}$ 
in \citet{NN} was reduced to 97.9~km~s$^{-1}$ due to the revised distance 
estimate.

\subsection{Spectral Features}
\label{sec:SpecFeatures}
An optical spectrum (6000--10000\AA) for almost every object in the 2MUPM 
sample was obtained in order to assign spectral types (Papers V, IX, and X). 
Most (132) spectra examined for this analysis were obtained for the 
classification of the 2MASS selected samples, but some spectra from other 
sources were used. Spectral references are given in Table~\ref{tab:Ha}. Of the 
152 2MUPM objects, there were 150 optical spectra available. Two dwarfs, 
2MASS~J01550354+0950003 and 2MASS~J14283132+5923354 were confirmed and 
spectral typed using infrared spectra and we have not yet obtained optical 
spectra. 

The \textit{splot} routine in IRAF was used to visually inspect each spectrum 
and measure H$\alpha$ equivalent width (EW) and line flux. H$\alpha$ emission 
was detected in 63 of 81 M dwarfs and 15 of 69 L dwarfs, and upper limits were 
placed on H$\alpha$ non-detections by measuring the EW of a representative 
noise spike. Because emission strength for mid- to late-L dwarfs can be as 
small as 1-2\AA, we consider objects with H$\alpha$ emission distinguishable 
from noise as active rather than establishing a cut-off between active and 
inactive. 

We combined H$\alpha$ line flux with bolometric flux to obtain 
$log(F_{H\alpha}/F_{bol})$. Bolometric fluxes were 
calculated for each object using 2MASS $K_S$ magnitudes and the bolometric 
correction (BC$_K$) for $K_S$ magnitudes found by \citet{Golimowski04}. BC$_K$ 
is calculated from the polynomial fit to a spectral type/bolometric magnitude 
relation. For two objects, we were able to obtain H$\alpha$ EW but not line 
flux because the spectra were not flux calibrated. H$\alpha$ measurements and 
upper limits are presented in Table~\ref{tab:Ha}, and the activity properties 
of the 2MUPM sample are discussed in detail in \S~\ref{sec:ActResults}.

For binary objects unresolved in 2MASS, the bolometric flux is calculated from 
the combined magnitude. This presents a problem when measuring H$\alpha$ 
emission from the combined spectra because we do not know what fraction of the 
total activity is emitted by which component. If one dwarf is active and the 
other inactive, the ratio of measured H$\alpha$ flux to the combined 
bolometric flux results in a smaller relative emission strength than would be 
calculated for the single active dwarf. Because we use the log of a ratio to 
indicate H$\alpha$ strength, the resulting uncertainty depends on the relative 
contributions of each component to the bolometric flux, which in turn depends 
on $\delta K_S$ (the difference of the magnitudes).

If both components have equal emission strengths and $K_S$ magnitudes, then 
the measured $log(F_{H\alpha}/F_{bol})$ will be correct. For an equal mass 
binary with one active and one inactive component, $log(F_{H\alpha}/F_{bol})$ 
is 0.30 dex lower than it would be for the single active object. The effect 
will be smaller for an unequal mass binary; for $\delta K_S=2$ and activity 
only from the more luminous object, our calculated $log(F_{H\alpha}/F_{bol})$ 
is 0.08 dex lower than it would be for the single active object. In 
Figures~\ref{fig:STHalpha} and~\ref{fig:VtanH}, binaries are plotted as shaded 
symbols with the maximum uncertainties resulting from the combined magnitudes 
(+0.30 dex). The LHS1070 system has resolved spectroscopy and there is no 
uncertainty as to which component is producing H$\alpha$ emission. We used 
relative photometry from \citet{Ratzka} to determine the bolometric flux of 
the B and C components, and measured H$\alpha$ from resolved spectroscopy 
(Leinert, priv. comm.).


The wavelength range of our optical spectra enabled examination of the 2MUPM 
sample for the \ion{Li}{1} absorption line. According to the ``lithium test,'' 
the detection of the \ion{Li}{1} absorption feature is a sufficient, but not 
necessary, indicator of substellar mass \citep{LiTest,Basri98}. We have 
detected \ion{Li}{1} in 3 of the 69 L dwarfs in the 2MUPM sample, which is a 
significantly smaller fraction than the one third found in the \citet{K99} 
sample of 25 L dwarfs. This disparity is likely due to the lower 
signal-to-noise ratio of our spectra (obtained mostly with 4-m telescopes), 
compared to the \citet{K99} Keck 10-m spectra, so the small number of 
\ion{Li}{1} detections is likely an observational effect rather than a 
physical property. Individual objects with \ion{Li}{1} detections are 
discussed in papers V, IX, and X but there are not sufficient detections to 
use \ion{Li}{1} as an age diagnostic in the 2MUPM sample.

For the objects that overlap with the \citet{NN} sample, comparison of our 
H$\alpha$ EW measurements yields some interesting differences. Of 21 objects, 
15 had comparable (within 3~\AA) H$\alpha$ EW measurements. Three of the six 
dwarfs with significantly different measurements  have been observed in flare 
and are discussed in \S~\ref{sec:FlareObjects}. The other three objects 
have differences of 6--15~\AA. It is likely that these differences are 
evidence of variability rather than the result of different signal-to-noise or 
measuring techniques. Measurements for these three objects which show smaller 
variations are listed with those of objects showing stronger variability in 
Table~\ref{tab:flare}. For objects observed in both flare and quiescence, the 
quiescent measurement is used for sample statistics. We discuss activity 
properties of the 2MUPM sample in \S~\ref{sec:ActResults} and variable 
activity in \S~\ref{sec:FlareObjects}.

\section{Kinematics}
\label{sec:KineResults}
\subsection{Overall Properties}
\label{sec:KineStat}
The photometric selection of the 2MUPM sample provides an opportunity to 
study kinematics without bias and derive age estimates from a well-defined 
sample. The 2MUPM sample is characterized by a mean tangential velocity of 
$\langle V_{tan} \rangle=31.5~km~s^{-1}$ and a velocity dispersion of 
$\sigma_{tan}=20.7~km~s^{-1}$. There are two outliers 
(V$_{tan}>100~km~s^{-1}$) excluded from the kinematic analysis; these fast 
moving dwarfs are discussed in \S~\ref{sec:FastObjects}. The mean velocity and 
velocity dispersion show good agreement with the kinematics of the \citet{NN} 
sample. 

To obtain a kinematic age estimate for the 2MUPM sample, we use $\sigma_{tan}$ 
to estimate the total velocity dispersion ($\sigma_{tot}$) with the equation 
$\sigma_{tot}=(3/2)^{1/2}\sigma_{tan}$. Velocity dispersion is converted to 
age using the relation found by \citet{AgeEst}
\begin{displaymath} \sigma_{tot}=(10 km/s)\times[1+t/\tau]^{1/3} 
\end{displaymath} 
where $\tau=2\times10^8~yrs$ and $t$ is mean population age in years. With 
$\sigma_{tan}=20.8~km~s^{-1}$ (which excludes the outliers discussed in 
\S~\ref{sec:FastObjects}), this relation yields an age estimate of 3.1 Gyr for 
the 2MUPM sample. This kinematic age is sensitive to the spectrophotometric 
distances used to calculate V$_{tan}$. Allowing for the possibility of a 
systematic 10\% over- or underestimation of the distances, we calculate that 
age estimates could vary from 2.2--4.2~Gyr. This is in excellent agreement 
with the kinematic age estimate of 2--4~Gyr found by \citet{Dahn02}.

We can use kinematics to further investigate the age distribution of M and L 
dwarfs. Monte Carlo simulations of the substellar mass function by 
\citet{Burgasser05} produce a modeled age distribution with respect to 
T$_{eff}$, shown in his Figure 8. Though there is a large spread of ages in 
each spectral type/temperature bin, the general trend shows an older mean age 
for late-M and early-L dwarfs and a younger mean age for late-L dwarfs. This 
is because the L spectral class encompasses a combination of stellar and 
substellar objects. Only younger, relatively warm brown dwarfs have T$_{eff}$ 
that corresponds to L spectral types. Stars can have spectral types as late as 
$\sim$L4 and are likely to be older than brown dwarfs with the same T$_{eff}$ 
\citep{BRev01}. One might expect that the combination of both younger brown 
dwarfs and older stars in late-M to early-L types would produce a wider 
velocity distribution while late-L types contain exclusively brown dwarfs and 
would have a narrower distribution.

\begin{figure}
\plotone{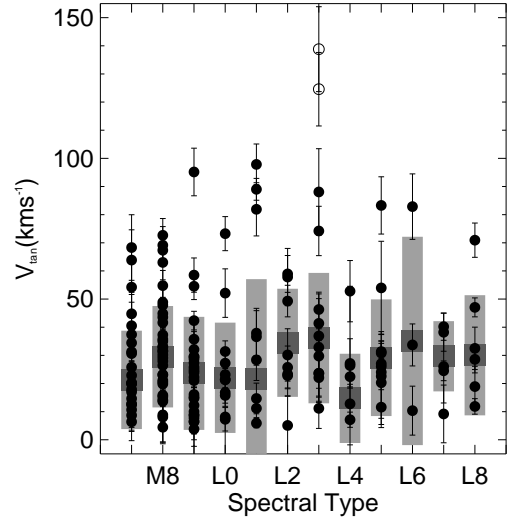}
\caption[Spectral type vs. V$_{tan}$]{Spectral type vs. V$_{tan}$. Half-integer
spectral types are rounded down. V$_{tan}$ for the outliers \textit{(open 
circles)} is distinguished from the data \textit{(filled circles)} that is 
used to calculate the mean and dispersion. The mean V$_{tan}$ 
\textit{(squares)} and its associated one sigma deviation 
\textit{(shaded bars)} are shown for each spectral type bin.}\label{fig:STVtan}
\end{figure}

To compare the expected ages with the kinematic distribution of the 2MUPM 
sample, we plot V$_{tan}$ as a function of spectral type 
(Figure~\ref{fig:STVtan}). We also plot the mean velocity and standard 
deviation for each spectral type bin to aid kinematic interpretation. Across 
M7 to L8 spectral types, the mean velocities per spectral type are largely 
constant. They are scattered within a range of 
$\langle V_{tan}\rangle=15~km~s^{-1}$ to 
$\langle V_{tan}\rangle=35~km~s^{-1}$, but there is no recognizable trend 
that hints at the expected age distribution.

\begin{figure}
\plotone{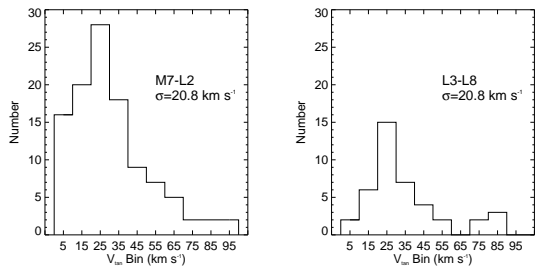}
\caption[Velocity Histograms] {Distribution of V$_{tan}$ for M7-L2 
\textit{(left)} and L3-L8 \textit{(right)} dwarfs. The center of each 
10~km~s$^{-1}$ wide bin is labeled. 
We find the dispersion ($\sigma$) to be 20.8~km~s$^{-1}$ for the M7-L2 portion 
of the sample, and 20.8~km~s$^{-1}$ for the L3-L8 portion. The histograms and 
the dispersions exclude the two outliers discussed in~\S~\ref{sec:FastObjects} 
which are both L3 dwarfs with V$_{tan} > 100~km~s^{-1}$.}\label{fig:VSThisto}
\end{figure}

Figure~\ref{fig:VSThisto} shows velocity distribution histograms of the M7--L2 
and L3--L8 populations. Both distributions peak at the bin centered on 
25~$km~s^{-1}$, and the velocity dispersions of the populations are nearly 
equal (M7--L2 with $\sigma_{tan}=20.8~km~s^{-1}$ and L3--L8 with 
$\sigma_{tan}=21.0~km~s^{-1}$), implying no age difference. It is possible 
that the 2MUPM sample is too small for kinematics to distinguish between stars 
and brown dwarfs, but it is also likely that the age effect is not pronounced 
enough to be apparent in kinematics.

\subsection{Kinematic Outliers}
\label{sec:FastObjects}
The three fastest objects in the 2MUPM sample warrant additional discussion. 
The fastest dwarf is \object{2MASSI J1721039+334415} (hereafter 2M1721+33), an
 L3 with a velocity of $V_{tan}=138.8\pm15.1~km~s^{-1}$; the next fastest is 
\object{2MASS J02511490-0352459} (hereafter 2M0251$-$03), an L3 with a 
velocity of $V_{tan}=124.6\pm13.1~km~s^{-1}$; and third fastest is 
\object{2MASSW J1300425+191235} (hereafter 2M1300+19), an L1 dwarf with a 
velocity of $V_{tan}=97.9\pm7.2~km~s^{-1}$. Both 2M0251$-$03 and 2M1721+33 are 
4$\sigma$ faster than the mean and are excluded from our kinematic analysis. 
While 2M1300+19 also has a high velocity, it is not as unusual as the two 
kinematic outliers and is included in our kinematic analysis.

In addition to their fast velocities, 2M1300+19 and 2M1721+33 are also 
unusually blue for their spectral types. Bluer colors suggest low metallicity 
and old dwarfs are likely to have high velocities. \citet{NN} discussed 
2M1300+19 because the combination of its unusually blue color and high 
velocity indicate that it is likely to be old. Our calculated 
$V_{tan}=97.9~km~s^{-1}$ is the third fastest velocity in the 2MUPM sample. 
Both 2M1200+19 and 2M1721+33 are discussed and spectra are presented in Papers 
V and IX because the combination of their slightly blue colors and faster 
kinematics suggest thick disk membership.

The remaining outlier, 2M0251$-$03, is also on the blue end of the color 
distribution but does not have unusual colors like 2M1300+19 and 2M1721+33. 
The spectrophotometric distance used to calculate its V$_{tan}$ is 
$d_{phot}=12.1\pm1.1~pc$, which is consistent with a preliminary distance of 
$d=12.7\pm1.2~pc$ found by the CTIOPI parallax program (Bartlett 2006, priv. 
comm.). Our measured proper motion of $\mu=2.17\pm0.11\arcsec~yr^{-1}$ at 
$PA=149\pm2\degr$ is consistent with the proper motion of 
$\mu=2.19\pm0.06\arcsec~yr^{-1}$ at $PA=149\degr$ measured by 
\citet{Deacon05}. 
Tangential velocity is only two of the three components of the total velocity, 
and a slow radial velocity would place 2M0251$-$03 closer to the mean of the 
kinematic distribution. While 2M0251$-$03 may be an unusual object, it is 
likely that its high velocity is simply the tail of the disk kinematic 
distribution.

\section{Activity}
\label{sec:ActResults}
\subsection{Activity and Spectral Type}
\label{sec:Act+ST}
\begin{figure}
\plotone{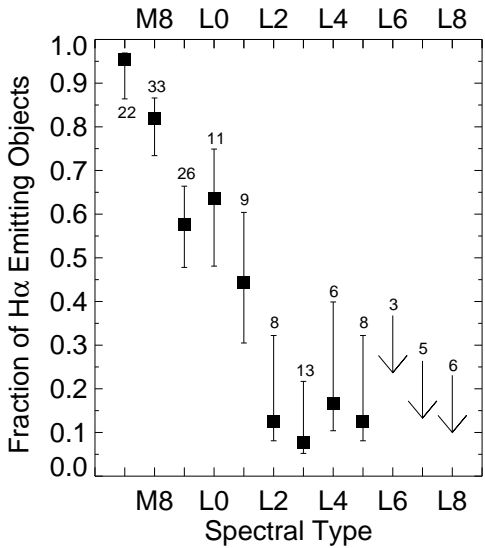}
\caption[Active Fraction vs. Spectral Type]{Fraction of objects with H$\alpha$ 
detection as a function of spectral type. The number above or below each data 
point is the total number of objects in each spectral type bin. The 
uncertainties are based on a binomial distribution. Upper limits are shown as 
down arrows for spectral type bins with no H$\alpha$ emitting 
objects.}\label{fig:histo}
\end{figure}

\begin{figure}
\plotone{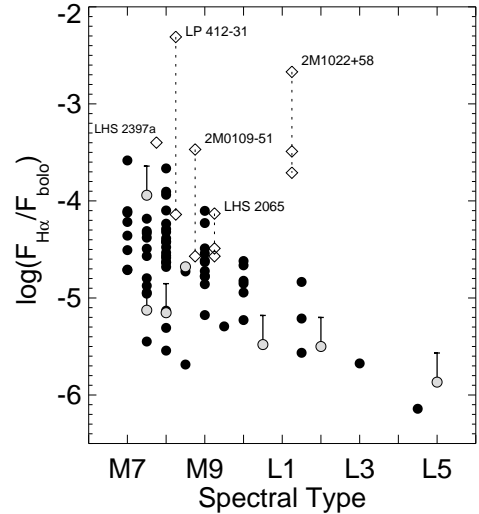}
\caption[Relative H$\alpha$ flux vs. Spectral Type]{Log of the ratio of 
H$\alpha$ flux to bolometric flux as a function of spectral type for objects 
with H$\alpha$ detections. Multiple systems \textit{(shaded circles)} are 
distinguished from single objects \textit{(filled circles)}. Multiple objects 
with unresolved spectroscopy are plotted with uncertainties that reflect the 
maximum uncertainty rising from the possibility that only one of the pair is 
active. Variable objects \textit{(diamonds)} are labeled just above the 
strongest observed H$\alpha$ flux. They are plotted offset slightly from their 
spectral type with flare and quiescent emission connected by a dashed line. We 
include LHS 2065 though we did not observe it in peak flux. We only observed 
LHS 2397a during a flare event, so there is no quiescent data 
point.}\label{fig:STHalpha}
\end{figure}

\begin{figure}
\plotone{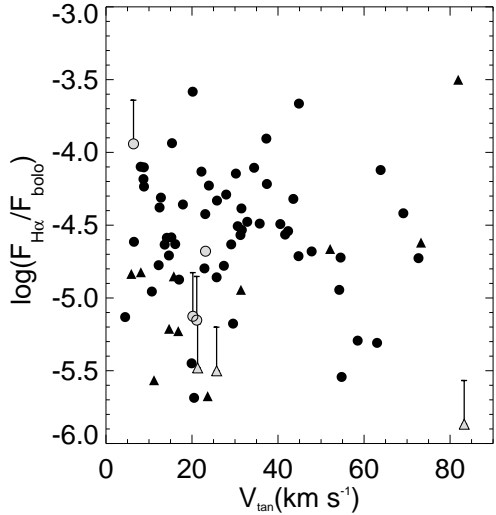}
\caption[Relative H$\alpha$ flux vs. V$_{tan}$] {Relative H$\alpha$ flux vs. 
V$_{tan}$ for active late-M dwarfs \textit{(circles)} and L dwarfs 
\textit{(triangles)} in the 2MUPM sample. Multiple systems \textit{(shaded)} 
are distinguished from single objects \textit{(black)}. Multiple objects 
with unresolved spectroscopy are plotted with uncertainties that reflect the 
maximum uncertainty rising from the possibility that only one of the pair is 
active. Data for LHS 2397a is not plotted because our spectrum was taken while 
it was in flare.}\label{fig:VtanH}
\end{figure}

Previous work has shown that the presence and strength of H$\alpha$ emission 
in late-M and L dwarfs decreases with lower mass and later spectral type.
Figure~\ref{fig:histo} plots the fraction of active objects per spectral type 
for the 2MUPM sample. M7 dwarfs have the largest activity fraction, with 21 of 
22 objects (95\%) showing H$\alpha$ emission. The activity fraction declines 
with later spectral type, but it does not go to zero. H$\alpha$ emission is 
present in 4 of 50 dwarfs between spectral types L2 and L8.

We find that the activity fraction of the 2MUPM sample drops rapidly through L 
dwarf sub-types. Half of L0 
dwarfs are active, one-fifth to two-fifths of L1 dwarfs, and approximately 
one-tenth for spectral types L2 and later. While there is no doubt that the 
activity fraction drops, it is possible that the steepness of the drop is due 
to small numbers and observational effects rather than the activity 
properties of L dwarfs. For spectral types L3 and later, H$\alpha$ was only 
detected in spectra taken with 8-m or 10-m telescopes. The EW of those 
detections is smaller than the upper limit of most spectra taken with smaller 
telescopes. \citet{NN} suggest that a lowered continuum surrounding the 
H$\alpha$ emission feature should make up for the decreased sensitivity, but 
to fully investigate the activity properties of late-L dwarfs, higher 
signal-to-noise spectra are needed. While there are only a few H$\alpha$ 
detections for types L2 to L8, activity has been detected in at least three T 
dwarfs \citep{Burgasser03}.

The peak activity fraction at M7 is consistent with previous results. 
\citet{NN} (Figure~6) found that all M7 and M8 dwarfs in their sample are 
active. Their activity fraction declines with later spectral type and no 
H$\alpha$ emission is found in types L5 and later. \citet{West04} (Figure~1) 
found that 73\% of M8 dwarfs in their sample show H$\alpha$ emission and the 
activity fraction similarly declines with later spectral type. They 
investigate the possibility that their maximum activity fraction is lower 
(73\% rather than 100\%) due to the Galactic distribution of their sample. 
The \citet{West04} sample, photometrically selected from SDSS with distances 
as large as 200~pc, includes a kinematically older portion of dwarfs due to 
its larger mean scale height above the Galactic disk. The distance limits of 
the 2MUPM sample and the \citet{NN} sample (20~pc and 25~pc) exclude dwarfs 
with large scale heights and thus include a younger population.

Though EW is a standard way to characterize emission and absorption features, 
it is not an accurate measure of activity strength across all ultracool 
spectral types. The continuum flux surrounding the H$\alpha$ feature decreases 
with later spectral type for M and L dwarfs, so the ratio of H$\alpha$ line 
flux to bolometric flux is used instead of EW to compare the strength of 
H$\alpha$ emission across a wide range of spectral types \citep{Reid95}. Our 
method for computing $log(F_{H\alpha}/F_{bol})$ is discussed 
in~\S~\ref{sec:SpecFeatures}.

Early-type M dwarfs, like solar-type stars, are generally believed to be 
powered by rotational dynamos, driven via the interface between the rotational 
core and the convective outer envelope \citep{Parker55}. M dwarfs become fully 
convective at spectral type $\sim$M3, and activity in later-type stars is 
attributed to a turbulent dynamo \citep{Durney93}. However, the chromospheric 
activity distribution gives no indication of this transition: the average 
level of activity for M1 to M5 dwarfs with detected H-alpha remains constant 
at $\langle log(F_{H\alpha}/F_{bol})\rangle=-3.7$ \citep{PMSU2,West04}. At 
spectral types beyond M6, the activity level diminishes sharply 
\citep{Martin99,NN,Burgasser02,Mohanty03,West04}.

Figure~\ref{fig:STHalpha} shows activity strength plotted as a function of 
spectral type for the 2MUPM sample. Known multiple systems with unresolved 
spectroscopy are plotted at the spectral type of the primary with shaded 
symbols. As discussed in \S~\ref{sec:SpecFeatures}, multiple systems (with the 
exception of LHS1070) are shown with an upper error bar that shows the maximum 
effect of using a combined magnitude to estimate bolometric flux. Taking this 
correction into account, the strength of emission from multiple systems 
appears to be the same as for single objects. The active B component of 
resolved binary LHS1070 is an M7.5 dwarf with a 
$log(F_{H\alpha}/F_{bol})=-4.67$, which does not fall above the mean for its 
spectral type. The binary systems in the 2MUPM sample have estimated 
separations too large (1-10 AU) and estimated masses too small ($<0.1M_\odot$) 
for any known binary interaction mechanism to enhance chromospheric activity 
\citep{Cuntz00,Burgasser06}. For the 2MUPM sample, binarity and activity are 
uncorrelated.

Activity strength shows a strong downward trend spanning spectral types 
M7--L5, which is consistent with previous results. This suggests a correlation
 between effective temperature and activity strength for this spectral type 
range. As discussed above, this downward trend begins at spectral type 
$\sim$M6, which is possible correlated with a break down of the 
rotation/activity relation for earlier-type dwarfs \citep{Basri01,HighRes}. 
\citet{Mohanty02} find that the decrease is likely due to the high electrical 
resistivities in the cool, mostly neutral atmospheres of these ultracool 
objects, but the topic is still open to investigation.

\subsection{Activity and Age}
\label{sec:Act+Age}
Solar-type main sequence stars exhibit a direct correlation between age and 
activity; this is generally attributed to rotational spin-down, and a 
consequent reduction in the energy output of the rotational dynamo. As noted 
in the previous section, mid- and late-type M dwarfs are fully convective, and 
are therefore not expected to show the same dependence between activity and 
rotation. Indeed, there is observational evidence that these late-type dwarfs 
exhibit different age-activity relationships \citep{NN,Bochanski05}. For the 
2MUPM sample, we find that there is a large scatter in the plot of H$\alpha$ 
emission strength ($log(F_{H\alpha}/F_{bol})$) as a function of V$_{tan}$ 
(Figure~\ref{fig:VtanH}). Velocity is not a precise age indicator, but a loose 
correlation would be expected if ultracool dwarfs followed a solar-type 
activity-age relation.

\begin{figure}
\plotone{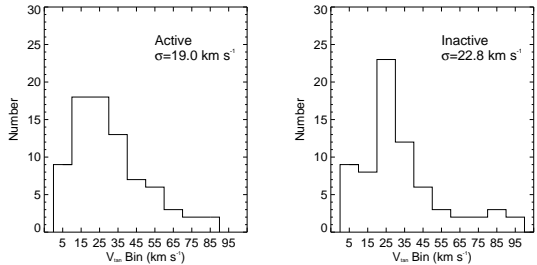}
\caption[Active and Inactive Velocity Histograms] {Distribution of V$_{tan}$ 
for dwarfs with \textit{(left)} and without \textit{(right)} H$\alpha$ 
emission. We find a dispersion ($\sigma$) of 19.0~km~s$^{-1}$ for active 
dwarfs and 22.8~km~s$^{-1}$ for inactive dwarfs. A K-S test indicates no 
significant difference between the two populations. The histograms and the 
dispersions exclude the two outliers discussed in~\S~\ref{sec:FastObjects} 
which are both inactive dwarfs with 
V$_{tan} > 100~km~s^{-1}$.}\label{fig:ActHisto}
\end{figure}

Previous work has shown that active M dwarfs, as a population, have younger 
kinematics than inactive M dwarfs \citep{Reid95,Bochanski05}. To further 
investigate that activity/age relationship for late-M and L dwarfs, we plot 
velocity distributions of the 2MUPM sample with and without H$\alpha$ emission 
(Figure~\ref{fig:ActHisto}). The active dwarfs have a smaller velocity 
dispersion ($\sigma_{tan}=19.0~km~s^{-1}$) and a slightly slower peak 
($\langle V_{tan} \rangle=29.9~km~s^{-1}$) than the inactive dwarfs 
($\sigma_{tan}=22.8~km~s^{-1}$, $\langle V_{tan} \rangle=33.3~km~s^{-1}$) 
which implies that the active dwarfs are younger than the inactive dwarfs. The 
velocity dispersions produce age estimates of $\sim$2 Gyr and $\sim$4 Gyr 
respectively. There is not a sufficient number of dwarfs, however, to conclude 
that the two populations have significantly different ages.

\citet{West06} show that observations of SDSS M7 dwarfs are consistent with a 
constant level of activity followed by a rapid decrease in emission at an age 
of 6--7 Gyr. Further investigation indicates that this activity lifetime is 
longer with later spectral type (West et al., in prep). If this describes 
activity in the 2MUPM sample, then our population is too young (2--4 Gyr) to 
show the effect of age on activity.

\begin{figure}
\plotone{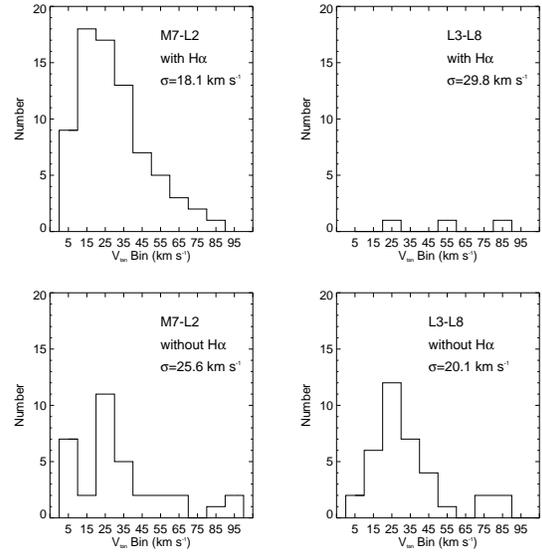}
\caption[Velocity Histograms of early and late spectral types with and without 
H$\alpha$ emission] {Distribution of V$_{tan}$ for M7--L2 dwarfs with 
H$\alpha$ emission \textit{(top left)}, M7--L2 dwarfs without H$\alpha$ 
emission \textit{(bottom left)}, L3--L8 dwarfs with H$\alpha$ emission 
\textit{(top right)} and L3--L8 dwarfs without H$\alpha$ emission 
\textit{(bottom right)}. The center of each 10~km~s$^{-1}$ wide bin is labeled 
and the $\sigma$ of each distribution is given on each histogram. The 
histograms and the dispersions exclude the two outliers discussed 
in~\S~\ref{sec:FastObjects} which are both inactive L3 dwarfs with 
V$_{tan} > 100~km~s^{-1}$.}\label{fig:Vhisto}
\end{figure}

To examine the kinematics in more detail, we plot the tangential velocity 
distributions of four different portions of the 2MUPM sample 
(Figure~\ref{fig:Vhisto}). These histograms divide spectral types M7-L2 
(likely stellar) from L3-L8 (likely substellar) and H$\alpha$ emitters from 
non-emitters. The active M7-L2 population shows both a smaller $\sigma_{tan}$ 
and a slower peak V$_{tan}$ than both of the inactive distributions. However, 
a K-S test between the distributions indicates that there is no significant 
difference between the populations. \citet{NN} suggested, based on \ion{Li}{1} 
detections, that active L dwarfs were drawn from an older, more massive 
population but the kinematics of the active L3--L8 dwarfs in our sample do not 
confirm this result. The active and inactive L dwarfs in the 2MUPM sample are 
kinematically indistinguishable, which is not surprising considering the small 
numbers of active L dwarfs. A larger sample would likely provide more insight 
into the age distribution of active L dwarfs.

\section{Variable Activity}
\label{sec:FlareResults}
\subsection{Variability in the 2MUPM Sample}
\label{sec:flarestat}
Out of the 150 objects in the 2MUPM sample with optical spectra, there are 11 
dwarfs that exhibit evidence of variability. Four of these have two 
observations with only 5--15~\AA~differences between the H$\alpha$ EW 
measurements. While their H$\alpha$ fluxes are not strong enough to classify 
them as flares, it is possible that this smaller scale variability is evidence 
that they are flaring objects not yet caught in peak flux. Both LHS 2065 and 
LP 412-31 have small scale variability as well as observed flare events. It is 
also possible that many more ultracool dwarfs have small scale variability but 
the effect has not yet been observed.

Seven dwarfs (5 M and 2 L) in the 2MUPM sample have been observed during flare 
events. Five (4 M and 1 L) were observed in flare as part of 
our observations of the 2MASS selected sample. If we assume that 
all objects in the 2MUPM sample are equally likely to flare, the number of 
objects observed in flare allows us to estimate that each dwarf spends 5/151 
or $\sim$3\% of their time in flare. If we divide the sample by spectral type, 
the late-M dwarfs in our sample have a flare rate of 4/81 ($\sim$5\%). 
\citet{Reid99} found a flare rate of 7\% through monitoring of BRI 0021, a 
M9.5 flare, and \citet{NN} similarly found a flare duty cycle of 7\% for 
M7--M9.5 dwarfs.

Using the same logic for L dwarfs, we find a flare rate of 1/69 ($\sim$1\%). 
This should be treated only as a lower limit due to the difficulty of 
observing weak H$\alpha$ in these faint objects. Many L dwarfs, especially 
spectral types L3 and later, only have one existing optical spectrum and not 
all those spectra have high enough signal-to-noise to observe H$\alpha$ at the 
expected strength for their spectral types (apprx. 1--2~\AA). For L dwarfs 
with observed activity and one epoch of data, we have assumed emission is 
quiescent rather than the product of a flare event. There is also a 
possibility that early-L dwarfs are more likely to flare than late-L dwarfs, 
just as they are more likely to have quiescent emission. More observations, 
including higher signal-to-noise optical spectra and monitoring for 
variability, are needed to understand the activity properties of L dwarfs.

\subsection{Variable Objects of Interest}
\label{sec:FlareObjects}
\begin{figure}
\plotone{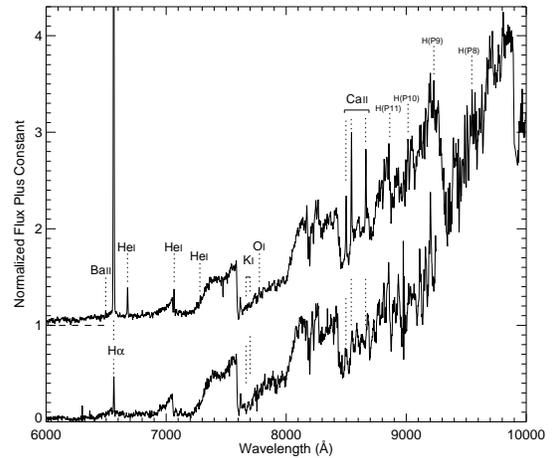}
\caption[Spectrum of LP412$-$31]{LP412$-$31, an M8 shown above in flare and 
below in quiescence. The dashed line marks the constant added to the flare 
spectrum. Various emission lines are labeled.}\label{fig:U10179}
\end{figure}

\begin{description}
\item[2MASS~J01443536$-$0716142] (hereafter 2M0144$-$07) is an L5 that 
was observed during a flare event by \citet{Liebert03} on 2001 Feb 20 with a 
maximum H$\alpha$ EW of 23~\AA~and on 2002 Jan 24 with no H$\alpha$ emission 
(upper limit 3~\AA). \citet{Liebert03} provide measurements of the H$\alpha$ 
as it declined from 24~\AA~to 6~\AA~over the course of 15 minutes. The  
spectrum of 2M0144$-$07 in flare shows no other emission lines, but if they 
are proportionally weaker than the H$\alpha$ emission, they could be hidden 
under the noise.

\item[LP~412$-$31] is a strongly flaring M8. We observed LP~412$-$31 
on 2000 Oct 20 with an H$\alpha$ EW of 24~\AA. On 2002 Jan 23 we observed it 
during a flare event with an H$\alpha$ EW of 330~\AA~and various other 
emission lines. These two spectra are plotted in Figure~\ref{fig:U10179}. 
Table~\ref{tab:flare} shows H$\alpha$ EW for a selection of other observations 
of LP~412$-$31, ranging from 18~\AA~to 83~\AA~\citep{HighRes,Basri01}. The 
peak observed H$\alpha$ emission of 330~\AA~is more than a factor of 10 
stronger than the weakest quiescent value. The recorded 83~\AA~emission could 
be part of a different flare event or it could be evidence of small-scale 
variability in addition to larger flare events. These variations have not been 
checked for periodicity. 

\item[LHS~2065] is an M9 that was observed in flare by \citet{Martin01} 
(spectrum shown in their Figure 1) with a peak emission of 261~\AA~on 1998 
Dec 12. LHS~2065 exhibits small scale variability (between 7~\AA~and 25~\AA), 
both according to the other H$\alpha$ measurements presented in 
\citet{Martin01} and our own measurements, listed in Table~\ref{tab:flare}. 
This variability accompanied by a larger flare event is similar to the 
emission from the M8 dwarf LP~412$-$31, but the flare spectra for LHS~2065 is 
different because it shows H$\alpha$ with few other emission lines. Redward of 
H$\alpha$, the only additional emission is from the \ion{He}{1} line, and it 
is only discernible in the epoch with the strongest H$\alpha$ emission.

\item[LHS 2243] \citet{NN} observed this M8 dwarf in flare with an 
H$\alpha$ EW of 44~\AA~compared to a previous observation of 1.3~\AA~by 
\citet{Martin94}. We observed LHS 2243 on 11 June 2003 with an H$\alpha$ EW of 
7~\AA.

\item[2MASSI~J1108307+683017] (hereafter 2M1108+68) is an L0.5 dwarf 
that we observed on 2004 Mar 3 with no H$\alpha$ emission (upper limit 1~\AA) 
but \citet{NN} observed it in 1999 Jun with an H$\alpha$ EW of 8~\AA. 
Photometric variability studies in the \textit{I} band have found that 
2M1108+68 is variable but not periodic \citep{Gelino02,Clarke02}. 
\citet{Gelino02} suggest that the variability it is due to variations in the 
clouds at the surface of the dwarf and not activity. It is possible that the 
variable H$\alpha$ emission and the \textit{I} band variations are both due to 
chromospheric activity, but more observations are needed to compare the 
photometric variations with the H$\alpha$ variability.

\item[LHS 2397a] is a spectroscopically unresolved binary system with an M8 
dwarf primary and an L7.5 dwarf secondary \citep{Freed03}. We observed 
LHS~2397 on 2002 Jan 23 with an H$\alpha$ EW of 78~\AA, which is significantly 
stronger than the \citet{Martin99} observation of 12~\AA~in 1998 Dec. 
Previously, \citet{Bessell91} observed strong variability in H$\alpha$ 
emission between two spectra taken 10 minutes apart in 1982. 

\item[2MASSW J1707183+643933] (hereafter 2M1707+64) is an M9 dwarf 
that we observed on 2003 Jul 7 
with an H$\alpha$ EW of 28.7~\AA. \citet{NN} also observed it in 1999 Jun with 
an EW of 9.8~\AA. \citet{RockenfellerPeriod06} examine the photometric 
variability of 2M1707+64 and find a period of $\sim$3.6 hr, and 
\citet{RockenfellerFlare06} discuss photometric observations of this object 
during a particularly bright flare. 
\end{description}

\subsection{Strong H$\alpha$ Emission from the L1 dwarf 
2MASS~J10224821+5825453}
\label{sec:20373}
\begin{figure}
\plotone{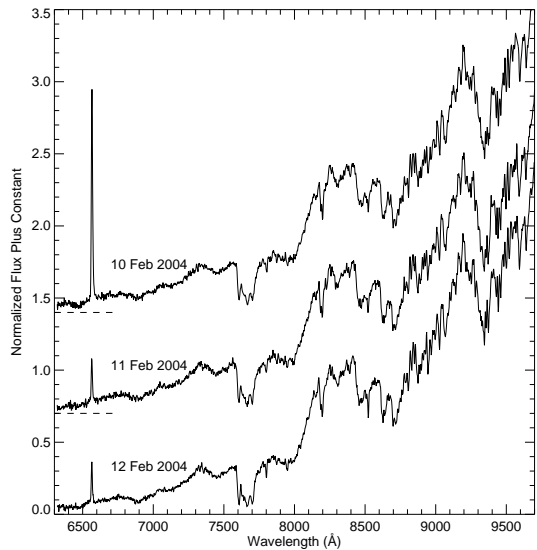}
\caption[Spectrum of 2M1022+58]{2M1022+58, an L1 on three successive 
nights observed with large H$\alpha$ flare of EW 128~\AA~on 10 Feb 2004 and 
EW of 24~\AA~and 26~\AA~on the next two nights. The dashed lines mark the 
constants added to the offset spectra.}\label{fig:U20373}
\end{figure}

We observed the L1 dwarf \object{2MASS J10224821+5825453} (hereafter 
2M1022+58) on 2004 Feb 10 with an H$\alpha$ EW of 128~\AA. The next 
two nights, 2004 Feb 11 and 12, the H$\alpha$ emission was at 24 and 26~\AA, 
respectively. Figure~\ref{fig:U20373} shows data from the three successive 
nights. The H$\alpha$ strength is enhanced by approximately an order of 
magnitude during the peak observed emission, but there are no other emission 
lines present. A lower signal-to-noise spectrum obtained on 2003 March 13 
shows no H$\alpha$ emission with an upper limit of $\sim$30~\AA. Without more 
observations, it is difficult to classify 2M1022+58 as a flaring dwarf or as a 
consistently strong, variable emitter.

On the plot of spectral type vs.~activity strength 
(Figure~\ref{fig:STHalpha}), 2M1022+58 is an outlier with remarkably strong 
quiescent emission of $log(F_{H\alpha}/F_{bol})=-3.5$ (peak emission 
$log(F_{H\alpha}/F_{bol})=-2.7$). It is possible that the 
$log(F_{H\alpha}/F_{bol})=-3.5$ emission is not the quiescent level but is due 
to small-scale variability or another large variation. This relatively strong 
quiescent emission is evidence that 2M1022+58 is not a flaring dwarf but is in 
a permanent state of strong chromospheric activity. 

There are three examples of consistently strong H$\alpha$ emission in the 
ultracool regime. The M9.5 star \object{PC 0025+0447} (hereafter PC~0025) has 
been monitored for several years and shows continually strong activity at 
$log(F_{H\alpha}/F_{bol}) \leq -3.4$ \citep{Schneider91,PC0025}. The limited 
observations of 2M1022+58 suggest similarities between the two objects, as 
H$\alpha$ emission in PC~0025 also varies by an order of magnitude. 
\citet{Hall02} suggest that the L3 dwarf \object{2MASSI J1315309-264951} 
(hereafter 2M1315$-$26) shows similar strong, variable activity. It has been 
observed twice with an H$\alpha$ EW of 
$\sim$120~\AA~($log(F_{H\alpha}/F_{bol})=-4$), and two other epochs of 
spectroscopy show H$\alpha$ EWs of 97~\AA~and 25~\AA. The third strong, 
variable dwarf is \object{2MASS J12373919+6526148} (hereafter 2M1237+65), a 
T6.5 dwarf monitored by \citet{Burgasser02} which consistently shows unusually 
strong emission at $log(F_{H\alpha}/F_{bol})=-4.3$. 

Aside from 2M1315$-$26 and 2M1022+58, there is only one other L dwarf that has 
been observed with variable activity. \citet{Liebert03} observed the L5 dwarf 
\object{2MASS J01443536-0716142} (hereafter 2M0144$-$07) during a flare event 
with maximum H$\alpha$ emission at $log(F_{H\alpha}/F_{bol})=-4.6$. A later 
observation shows no H$\alpha$ emission. One unexpected similarity between 
these three variable L dwarfs is the absence of other emission lines that 
usually accompany variable H$\alpha$ emission. For 2M0144$-$07 and 
2M1315$-$26, it is possible that other emission lines are present but simply 
below the noise. Emission lines in 2M1022+58, however, would be visible 
against the smooth continuum of the relatively high signal-to-noise spectra. 

The H$\alpha$ emission in 2M1022+58 increases by a factor of 10 in our 
observations. When PC~0025 undergoes a similar increase, it is always 
accompanied by the appearance of some other emission lines, most notably 
\ion{He}{1}. As an L1 with peak H$\alpha$ emission at least as strong as the 
emission of PC~0025, it is unknown why 2M1022+58 shows no other emission lines.

The mechanism responsible for H$\alpha$ emission in 2M1022+58 warrants further 
investigation. \citet{Burgasser02} and \citet{Martin99} investigated the 
possibility of a tight, low mass companion object causing the variable, strong 
H$\alpha$ emission from 2M1237+65 and PC~0025 respectively. Neither 
investigation produced significant evidence of a companion object. 

\citet{Liebert03} hypothesized that H$\alpha$ emission from 2M0144$-$07 could 
be the result of an active dynamo in a relatively young, low mass object. 
Despite a fast tangential velocity of V$_{tan}=81.9~km~s^{-1}$, this L1 dwarf 
has optical spectral features that indicate youth (details are in Paper X). 
While the high V$_{tan}$ suggests older age, it is based on only two of three 
velocity components. Kinematics are a good statistical age indicator but less 
accurate with single objects. For example, CM Draconis is a near-solar 
metallicity dwarf with a space velocity of 160~km~s$^-1$ 
\citep{Legget98,Viti02}. Further high signal-to-noise spectra should be 
obtained of this object, to characterize its variability, obtain a radial 
velocity, and determine whether significant lithium absorption is present.

\subsection{An Amazing Flare Event on 2MASS J1028404$-$143843}
\label{sec:10910}
\begin{figure}
\plotone{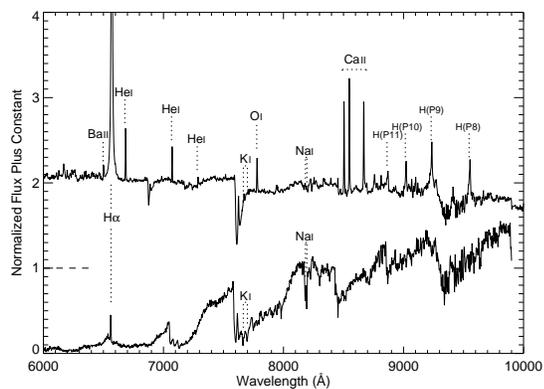}
\caption[Spectrum of 2M1028$-$14]{Spectacularly flaring 2M1028$-$14, an M7, 
shown above in flare and below in relative quiescence. The dashed line marks 
the constant added to the flare spectrum. Various emission lines are 
labeled.}\label{fig:U10910}
\end{figure}

During spectroscopic observations of the 2MASS selected sample, the M7 dwarf 
2MASS J1028404$-$143843 (hereafter 2M1028$-$14) was observed during an amazing 
flare event; the flare spectrum and a comparison quiescent spectrum are 
plotted in Figure~\ref{fig:U10910}. On 2002 Jan 25 2M1028$-$14 was in flare 
with many emission lines, including H$\alpha$ with an EW of 97~\AA, which is 
below 100~\AA~only because of the strongly enhanced flaring continuum. The EW 
and line fluxes for the flare observation are given in Table~\ref{tab:U10910}. 
Observations of 2M1028$-$14 in quiescence were taken on 2003 Mar 19 with an 
H$\alpha$ EW of 23~\AA~and on 2003 Mar 21 with an EW of 12~\AA. 

The 2M1028$-$14 flare event elevates the continuum well into the red optical 
spectra presented (6000~\AA~to $\sim$10000~\AA), to such an extent that the 
slope becomes blue. This effect is also found in the flare of 
\object{2MASSI J0149089+295613} (hereafter 2M0149+29), an M9.5 star described 
in \citet{Liebert99}. Observed over a similar wavelength region, the latter 
object also had an elevated continuum which extended from 6000~\AA~to 
$\sim$7600~\AA. Longward of 7600~\AA, however, the photospheric molecular 
absorption of 2M0149+29 appeared relatively similar in strength to that in 
quiescence. The elevated continuum extends farther for 2M1028$-$14, where all 
but the strong 9400~\AA~H$_2$O band appear completely masked.

The emission spectra of both flaring objects contain lines varying in 
excitation from \ion{Ba}{2} and \ion{He}{1} to the neutral alkalis \ion{Na}{1} 
and \ion{K}{1}, which are typically in absorption but have been thrown into 
emission. Table~\ref{tab:U10910} lists all the emission lines found in the 
flare spectrum. When scaled proportional to the H$\alpha$ feature, the 
emission spectrum of 2M1028$-$14 is weaker than that of 2M1049+29, except for 
the \ion{Ca}{2} triplet.

It is difficult to compare the strength of the 2M1028$-$14 flare with the most 
active M dwarfs since the emphasis of previous studies has been on the 
ultraviolet and blue spectra plus H$\alpha$. In these flare events, the 
erupting plasma reaches temperatures of order 10$^4$~K and forms a thermal 
bremsstrahlung continuum peaking in the far-ultraviolet 
\citep{Giampapa83,Eason92}. Coordinated studies in the far-ultraviolet used 
the International Ultraviolet Explorer satellite with ground-based telescopes 
to measure this continuum component in the far-ultraviolet for flares in 
YZ CMi, Proxima Cen, and AD Leo \citep{Foing86}. One of the most impressive of 
these coordinated studies was that of \citet{Hawley91} on ``great flare of 
1985 April 12 on AD Leonis.'' These studies, however, provide no red optical 
spectra for direct comparison to 2M1028$-$14.

\citet{Liebert99} argued that, if the ultraviolet flux scaled with H$\alpha$ 
in the same manner as it does in the M flare stars studied from space, 
2M0149+29 emitted a flare luminosity that exceeded the normal bolometric 
photospheric luminosity at the flare's peak. The same argument would apply to 
2M1028$-$14, but the greater relative strength of the red continuum flux 
suggests that the flare's peak luminosity could have exceeded that of the 
quiet photosphere even more than in the 2M0149+29 flare event.

We do not know whether the strongest M flare stars have red continua that turn 
blue like the 2M1028$-$14 event because we are not aware of an observation 
where a flare continuum was studied at these wavelengths. More observations of 
these known flare stars in the optical would provide direct comparison, as 
would ultraviolet observations of 2M1028$-$14 in flare.

\section{Summary}
\label{sec:conclusion}
We have presented proper motions, tangential velocities, H$\alpha$ 
measurements, and flare observations for a nearly complete, volume limited 
sample of 152 late-M and L dwarfs. We summarize our results:
\begin{itemize}
\item The 2MUPM sample has a mean tangential velocity of 
$\langle V_{tan}\rangle =31.5~km~s^{-1}$, and a tangential velocity dispersion 
of $\sigma_{tan} = 20.7~km~s^{-1}$. With the exception of two outliers, the 
distribution of tangential velocities is consistent with thin disk kinematics.
\item Velocity distributions of the 2MUPM sample are not sufficiently 
sensitive to show the expected age distribution of stars and brown dwarfs. 
Examination of the \ion{Li}{1} absorption feature is needed to further 
investigate the age distributions of stellar and sub-stellar populations.
\item We find that there is no direct correlation between activity and age in 
the 2MUPM sample. Velocity distributions show that the active population is 
marginally older than the inactive population, but more data are needed to 
draw kinematic conclusions.
\item In the 2MUPM sample, activity strength declines with later spectral type 
for types M7 and later; the correlation extends through L5 dwarfs. 
\item In agreement with previous results, we find the activity fraction to 
peak at spectral type M7 and decline through mid-L dwarfs for the 2MUPM 
sample. For spectral types L3 and later, the activity fraction is unclear due 
to the difficulty of obtaining high signal-to-noise spectra of these faint 
objects. More observations of L dwarfs are needed to understand the activity 
properties of these low mass objects.
\item The late-M dwarfs in our sample have an estimated flare rate of 
$\sim$5\% which is consistent with previous results. We put a lower limit of 
$\sim$2\% on the flare rate of L dwarfs, but more data is needed to obtain a 
more accurate estimate and investigate the occurrence of flares across all L 
dwarf subtypes. It is possible, for example, that early-L dwarfs flare more 
often that late-L dwarfs.
\item The L1 dwarf 2M1022+58 shows unexpectedly strong H$\alpha$ emission both 
in flare and quiescence. The absence of other emission lines in the flare 
spectrum remains unclear.
\item The M7 dwarf 2M1028$-$14 shows an amazing flare spectrum with many 
emission lines.
\end{itemize}

\acknowledgments
We thank Adam Burgasser and Sebastian Lepine for useful discussions.
S.~J.~S. is partially supported by a grant from the New York NASA Space Grant Consortium, by NASA through an award issued by JPL/Caltech.
This research was partially supported by a grant from the NASA/NSF NStars initiative, administered by JPL, Pasadena, CA.
K.~L.~C. is supported by an NSF Astronomy and Astrophysics Postdoctoral Fellowship under award AST-0401418.
Based on observations obtained at the Gemini Observatory, which is operated by the Association of Universities for Research in Astronomy, Inc., under a cooperative agreement with the NSF on behalf of the Gemini partnership: the National Science Foundation (United States), the Particle Physics and Astronomy Research Council (United Kingdom), the National Research Council (Canada), CONICYT (Chile), the Australian Research Council (Australia), CNPq (Brazil) and CONICET (Argentina).
We gratefully acknowledge the team that built CPAPIR: Etienne Artigau, Rene Doyon and Daniel Nadeau. We also thank Etienne Artigau and Rene Doyon for continuing support of the instrument. CPAPIR was built through grants from the Canadian Foundation for Innovation and the Natural Science and Engineering Research Council of Canada.
We thank the Research Consortium on Nearby Stars (RECONS) for a preliminary distance measured as part of the Cerro Tololo Inter-american Observatory Parallax Investigation (CTIOPI); CTIOPI was a National Optical Astronomy Observatory (NOAO) Survey Program and continues as part of the Small and Moderate Aperture Telescope Research System (SMARTS) Consortium. 
This publication makes use of data products from the Two Micron All Sky Survey, which is a joint project of the University of Massachusetts and Infrared Processing and Analysis Center/California Institute of Technology, funded by the National Aeronautics and Space Administration and the National Science Foundation; the NASA/IPAC Infrared Science Archive, which is operated by the Jet Propulsion Laboratory/California Institute of Technology, under contract with the National Aeronautics and Space Administration.
The Digitized Sky Survey was produced at the Space Telescope Science Institute under U.S. Government grant NAG W-2166. The images of these surveys are based on photographic data obtained using the Oschin Schmidt Telescope on Palomar Mountain and the UK Schmidt Telescope. The plates were processed into the present compressed digital form with the permission of these institutions.
The National Geographic Society - Palomar Observatory Sky Atlas (POSS-I) was made by the California Institute of Technology with grants from the National Geographic Society.
The Second Palomar Observatory Sky Survey (POSS-II) was made by the California Institute of Technology with funds from the National Science Foundation, the National Aeronautics and Space Administration, the National Geographic Society, the Sloan Foundation, the Samuel Oschin Foundation, and the Eastman Kodak Corporation.
The Oschin Schmidt Telescope is operated by the California Institute of Technology and Palomar Observatory.
The UK Schmidt Telescope was operated by the Royal Observatory Edinburgh, with funding from the UK Science and Engineering Research Council (later the UK Particle Physics and Astronomy Research Council), until 1988 June, and thereafter by the Anglo-Australian Observatory. The blue plates of the southern Sky Atlas and its Equatorial Extension (together known as the SERC-J), as well as the Equatorial Red (ER), and the Second Epoch [red] Survey (SES) were all taken with the UK Schmidt.
Supplemental funding for sky-survey work at the ST ScI is provided by the European Southern Observatory. 
We accessed the Digital Sky as Guest User, Canadian Astronomy Data Centre, which is operated by the Herzberg Institute of Astrophysics, National Research Council of Canada.



\clearpage

\begin{deluxetable}{llllllllll} \tablewidth{0pt}
\tablecaption{Kinematics of the 2MUPM Sample \label{tab:all}}
\rotate \tabletypesize{\scriptsize}
\tablehead{\colhead{2MASS Designation} & \colhead{UCD \#} & 
\colhead{Other Name} & \colhead{ST} & \colhead{d (pc)} & 
\colhead{$\mu_{tot}$ (\arcsec~yr$^{-1}$)} & \colhead{PA ($^\circ$)} & 
\colhead{V$_{tan}$ (km~s$^{-1}$)} & \colhead{ST Ref.} & \colhead{$\pi$ Ref.}}
\startdata
00043484$-$4044058\tablenotemark{a} &  20004& LHS 102B/GJ 1001B         & L5     &  9.55$\pm$ 1.04&  1.83$\pm$   0.10&  158$\pm$      3&   83.3$\pm$  10.2&      1&     1\\
0019262+461407    &  10013& \nodata                   & M8     & 19.50$\pm$ 1.60&  0.17$\pm$   0.06&  124$\pm$     15&   15.9$\pm$   5.7&      2&      \\
0019457+521317    &  10014& \nodata                   & M9     & 18.70$\pm$ 1.20&  0.29$\pm$   0.20&  111$\pm$     47&   25.7$\pm$  17.5&      2&      \\
0024246$-$015819  &  10018& BRI 0021-0214             & M9.5   & 11.55$\pm$ 0.53&  0.14$\pm$   0.09&  338$\pm$     41&    7.7$\pm$   5.1&      3&     2\\
0024442$-$270825  &  10019& LHS 1070b                 & M8.5   &  7.71$\pm$ 0.14&  0.63$\pm$   0.04&  348$\pm$      5&   23.1$\pm$   1.5&      4& 1,3  \\
                  &       & LHS 1070c                 & L0     &                &                  &                 &                  &       &      \\
0027559+221932\tablenotemark{a}    &  10022& LP 349- 25                & M8:    & 10.30$\pm$ 1.70&  0.43$\pm$   0.05&  113$\pm$      5&   21.1$\pm$   4.1&      5&      \\
00315477+0649463  &  20024& LSPM J0031+06             & M9     & 18.66$\pm$ 1.22&  0.28$\pm$   0.10&   76$\pm$     11&   24.7$\pm$   8.8&      6&      \\
\enddata
\tablecomments{Reference for $\pi$ is only listed if the distance given is a 
parallax measurement.}
\tablenotetext{a}{Binary considered as a single object due to the lack of resolved spectroscopy.} 
\tablenotetext{b}{Proper motion from \citet{Vrba04}}
\tablenotetext{c}{Wide binary} 
\tablerefs{Spectral Type References: 
(1) Kirkpatrick et al. 2000
(2) Paper 5
(3) Irwin et al. 1991
(4) Leinert et al. 2000
(5) Gizis et al. 2000
(6) Paper 10
(7) Phan-Bao et al. 2001
(8) Geballe et al. 2002
(9) Cruz \& Reid 2002
(10) Paper 9
(11) Delfosse et al. 1997
(12) Kirkpatrick et al. 1997
(13) Mart{\'{\i}}n et al. 1999b
(14) Kirkpatrick et al. in prep
(15) Kirkpatrick et al. 1995
(16) Hawley et al. 2002
(17) Thorstensen \& Kirkpatrick 2003
(18) Ruiz et al. 1991
(19) Gizis \& Reid 1997
(20) Henry et al. 2004
(21) Kirkpatrick et al. 1991
(22) Delfosse et al. 2001
(23) Gizis 2002
(24) Fan et al. 2000
(25) Ruiz et al. 1997
(26) Kirkpatrick et al. 1999
(27) Reid et al. 2003
(28) Phan-Bao et al. 2003
(29) Kirkpatrick et al. 2001}
\tablerefs{$\pi$ References:
(1) van Altena et al. 1995
(2) Tinney, Reid, Gizis, \& Mould 1995
(3) Costa et al. 2005
(4) Dahn et al. 2002
(5) Tinney 1996
(6) Vrba et al. 2004
(7) Thorstensen \& Kirkpatrick 2003
(8) Ianna \& Fredrick 1995
(9) Monet et al. 1992
(10) Deacon \& Hambly 2001
(11) Jao et al. 2005
(12) Reid et al. 2003
(13) ESA 1997
}
\end{deluxetable}

\clearpage

\pagestyle{empty}

\begin{deluxetable}{llllllllllll} \tablewidth{0pt}
\tablecaption{H$\alpha$ Emission Data for the 2MUPM Sample \label{tab:Ha}}
\rotate \tabletypesize{\scriptsize}
\tablehead{\colhead{2MASS Designation} & \colhead{UCD \#} &
\colhead{Other Name} & \colhead{ST} & \colhead{K} & 
\colhead{M$_{bol}$} & \colhead{F$_{bol}$} & \colhead{H$\alpha$ EW (\AA)} & 
\colhead{H$\alpha$ Flux} & \colhead{log($F_{H\alpha}/F_{bol}$)} & 
\colhead{UT date} & \colhead{Spectra Ref}}
\startdata
00043484$-$4044058\tablenotemark{a} & 20004& LHS 102B/GJ 1001B         & L5           &   11.4&  14.74& 3.58E-11& -1.50         & 4.86E-17&        -5.87& \nodata    &           \\       
0019262+461407    & 10013& \nodata                   & M8           &  11.47&  14.57& 4.18E-11& $>$ -4.55     & \nodata & \nodata     & 2002 Jul 8 &           \\       
0019457+521317    & 10014& \nodata                   & M9           &  11.62&  14.76& 3.50E-11& -9.25         & 4.85E-16&        -4.86& 2000 Oct 2 &           \\       
0024246$-$015819  & 10018& BRI 0021-0214             & M9.5         &  10.58&  13.76& 8.85E-11& $>$ -0.36     & \nodata & \nodata     & 2003 Jul 10&           \\       
0024442$-$270825  & 10019& LHS 1070b                 & M8.5         &   9.83&  12.95& 1.79E-10& -8.43         & 3.75E-15&        -4.68& \nodata    &          1\\       
                  &      & LHS 1070c                 & L0           &  10.14&  13.34& 1.25E-10& $>$ -0.38     & \nodata & \nodata     & \nodata    &          1\\       
0027559+221932\tablenotemark{a}    & 10022& LP 349- 25                & M8:          &   9.56&  12.66& 2.44E-10& -12.58        & 1.72E-15&        -5.15& \nodata    &           \\       
00315477+0649463  & 20024& LSPM J0031+06             & M9           &  11.72&  14.87& 3.18E-11& $>$ -3.75     & \nodata & \nodata     & 2003 Oct 12&           \\
\enddata
\tablecomments{Spectra references are only given when the data used to 
meaure H$\alpha$ was not taken as part of the 2MASS sample spectral typing.}
\tablenotemark{a} \tablenotetext{a}{Binary considered as a single object due to the lack of resolved spectroscopy.}
\tablenotetext{b}{Wide binary} 
\tablerefs{
(1) Leinert, priv. comm.
(2) Hawley et al. 2002
(3) Kirkpatrick, priv. comm.
(4) Kirkpatrick et al. 1999 
(5) Kirkpatrick et al. 1997
(6) Kirkpatrick et al. in prep
(7) Fan et al. 2000
(8) Kirkpatrick et al. 1995
(9) Kirkpatrick et al. 2000 
}
\end{deluxetable}

\clearpage

\pagestyle{plaintop}

\begin{deluxetable}{lllllll} \tablewidth{0pt}
\tabletypesize{\footnotesize}
\tablecaption{H$\alpha$ EW for Variable Objects \label{tab:flare}}
\tablehead{\colhead{2MASS} & \colhead{Other} & \colhead{2MUPM} & 
\colhead{Spectral} & \colhead{H$\alpha$} & \colhead{UT Date} & 
\colhead{Reference}\\ \colhead{Designation} & \colhead{Name} &
\colhead{Sample} & \colhead{Type} & \colhead{EW (\AA)} & 
\colhead{} & \colhead{}}
\startdata
01090150$-$5100494 & \nodata & yes & M9 & -114 & 2003 Nov 9 &  \\
 &  &  &  & -31 & 2004 Aug 9 &  \\
0144353$-$071614 & \nodata & yes & L5 & -24 & 2001 Feb 20 & 1 \\
 &  &  &  & $>$-3 & 2002 Jan 24 &  \\
0320596+185423 & LP 412$-$31 & yes & M8 & -29 & 1995 Oct 28 & 2 \\
 &  &  &  & -29 & 1998 Dec & 3 \\
 &  &  &  & -24 & 2000 Oct 1 &  \\
 &  &  &  & -330 & 2002 Jan 23 &  \\
\enddata
\tablerefs{
(1) \citet{Liebert03}
(2) \citet{Martin96}
(3) \citet{NN}
(4) \citet{HighRes}
(5) \citet{Basri01}
(6) \citet{Martin01}
(7) \citet{Martin94}
(8) \citet{Martin99}}

\end{deluxetable}

\clearpage

\begin{deluxetable}{lllll} \tablewidth{0pt}
\tablecaption{Line Strengths for 2M 1028$-$14 \label{tab:U10910}}
\tablehead{\colhead{Ion} & \colhead{$\lambda$ (\AA)} 
& \colhead{Continuum\tablenotemark{a}} & \colhead{Line Flux\tablenotemark{a}} 
& \colhead{EW (\AA)}}
\startdata
\ion{Ba}{2} & 6497 & 78 & 58 & -0.736 \\
H$\alpha$ & 6563 & 80 & 7745 & -96.66 \\
\ion{He}{1} & 6678 & 75 & 265 & -3.535 \\ 
\ion{He}{1} & 7067 & 74 & 158 & -2.146 \\ 
\ion{He}{1} & 7281 & 72 & 28 & -0.394 \\ 
\ion{O}{1} & 7777 & 66 & 180 & -2.727 \\
\ion{Ca}{2} & 8498 & 67 & 404 & -6.019 \\
\ion{Ca}{2} & 8542 & 68 & 548 & -8.124 \\ 
\ion{Ca}{2} & 8662 & 72 & 409 & -5.654 \\ 
H(P11) & 8863 & 72 & 100 & -1.395 \\
H(P10) & 9015 & 69 & 178 & -2.586 \\ 
H(P9) & 9230 & 80 & 204 & -2.55 \\ 
H(P8) & 9546 & 63 & 340 & -5.338 \\
\enddata
\tablenotetext{a}{in units of $10^{-16}~ergs~cm^{-2}~s^{-1}$}
\end{deluxetable}
\end{document}